\documentclass[aps,prb,twocolumn,superscriptaddress]{revtex4-1}

\usepackage{xcolor}
\usepackage{soul}
\usepackage{enumitem}
\usepackage{multirow}
\usepackage{amsmath}
\usepackage{graphicx}
\usepackage{color}
\usepackage{combelow}

\usepackage[normalem]{ulem}

\newcommand{\ie}{i.\,e., }
\newcommand{\eg}{e.\,g., }

\newcommand* \dzq{\ensuremath{\text{d}_{z^2}}}
\newcommand* \dxq{\ensuremath{\text{d}_{x^2-y^2}}}
\newcommand* \dxz{\ensuremath{\text{d}_{xz}}}
\newcommand* \dyz{\ensuremath{\text{d}_{yz}}}
\newcommand* \dxy{\ensuremath{\text{d}_{xy}}}

\newcommand{\out}[1]{}

\newcommand{\eref}[1]{Eq.~\ref{#1}}

\begin{document}

\title{Designing a mechanically driven spin-crossover molecular switch\\ via organic embedding}

\author{Sumanta Bhandary}
\affiliation{School of Physics, Trinity College Dublin, The University of Dublin, Dublin 2, Ireland}
\author{Jan M.\ Tomczak}
\affiliation
{Institute of Solid State Physics, Vienna University of Technology, 1040 Vienna, Austria}
\author{Angelo Valli}
\affiliation{Institute for Theoretical Physics, Vienna University of Technology, 1040 Vienna, Austria}

\begin{abstract} 
Among spin-crossover complexes, Fe-porphyrin (FeP) stands out for molecular spintronic applications: An intricate, yet favourable balance between ligand fields, charge transfer, and the Coulomb interaction makes FeP highly manipulable, while its planar structure facilitates device integration. Here, we theoretically design a mechanical spin-switch device in which external strain triggers the intrinsic magneto-structural coupling of FeP through a purely organic embedding. Exploiting the chemical compatibility and stretchability of graphene nanoribbon electrodes, we overcome common reliability and reproducibility issues of conventional inorganic setups. The competition between the Coulomb interaction and distortion-induced changes in ligand fields requires methodologies beyond the state-of-the-art: Combining density functional theory with many-body techniques, we demonstrate experimentally feasible tensile strain to trigger a low-spin ($S=1$) to high-spin ($S=2$) crossover. Concomitantly, the current through the device toggles by over an order of magnitude, adding a fully planar mechanical current-switch unit to the panoply of molecular spintronics.
\end{abstract}

\maketitle

The idea of utilising the electrons' spin for information-processing in molecular-scale devices is the foundation for the emerging field of {\it molecular spintronics}.~\cite{spintronics_bogani, spintronics_rocha,spintronics_sanvito,Schmaus2011} 
Elementary molecular devices, including spin-switches,~\cite{switch_ormaza, switch_sco} 
spin transistors,~\cite{zant10, kondo_transistor, trans_exch, trans_mae} or spin-valves,~\cite{valve}
exploit reversible control over magnetic states via 
spin-crossover, magnetic coupling, Kondo resonance, or magnetic anisotropy mechanisms.  
%
%
Among these, the molecular spin-crossover (SCO) is the most robust and versatile mechanism: It can be triggered upon contact with surfaces,~\cite{sco_gold_voltage, kuch2, kuch1} 
through external stimuli, \eg voltage~\cite{memristor, sco_gold_voltage} or strain,~\cite{frisenda16} and, importantly, it can operate reliably also at ambient conditions.~\cite{rt_sco} 
%

In standard experimental setups, a functional molecule is contacted between metallic electrodes (typically Au). However, the resulting properties strongly depend on the local atomic configuration of the junction, which is extremely difficult to control. 
Moreover, high atomic mobility and strong metal-molecule coupling often destroy intrinsic properties of the embedded molecule,~\cite{jia13,memristor} rendering its functionality unexploitable. 
To date, 
ensuring reliability and reproducibility remains a key challenge for device fabrication. 
A promising route to overcome the above issues is to integrate the magnetic molecule 
into a more compatible {\it organic} environment. 
In fact, carbon-based architectures have recently been demonstrated to yield highly mechanically stable 
and atomically precise molecular junctions,~\cite{cao12,jia15,he17,Li2019,li20} 
while at the same time preserving the intrinsic magnetic properties of the contacted molecule. 
Therefore it is timely to conceptualize functional molecular devices 
with organic embedding.

In this letter, we explore the theoretical design of a mechanically-driven SCO device, that consists of a magnetic iron-porphyrin (FeP) molecule 
covalently bonded to graphene nanoribbon (GNR) electrodes.
We identify a strategy to reversibly manipulate the molecular spin state by applying uniaxial strain. 
Exploiting the stretchability of the GNRs, we can directly engage the Fe-N coupling  that controls the competition between ligand-fields and the Coulomb interaction, at the heart of the SCO mechanism.

\begin{figure}[t]
\includegraphics[width=0.5\textwidth]{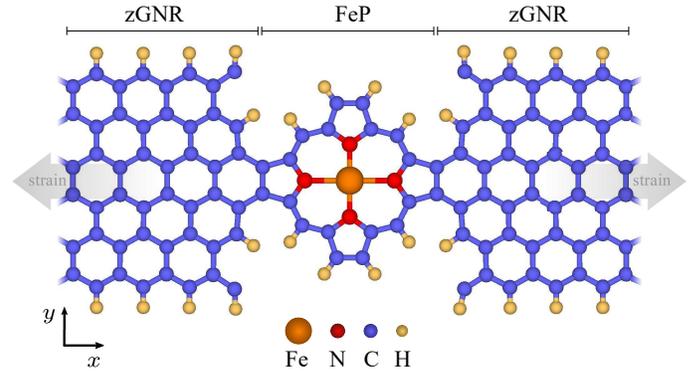}
\caption{\label{fig:unit_cell} Unit cell of the molecular device, in which zGNRs are chemically bonded to the FeP bridge. 
The spin state of the Fe atom is reversibly controlled by strain applied along the $x$-direction (grey arrows). }
\end{figure}

%
In Fig.~\ref{fig:unit_cell}, we present the device set-up where an FeP molecule is chemically anchored 
to semi-infinite zigzag-edge graphene nano-ribbon (zGNR) electrodes. 
The dangling bonds of the edge C atoms of the zGNRs are passivated with H atoms for an enhanced stability and, 
at the same time, preserving the $sp^2$ planar structure.~\cite{gnr_sum} 
The FeP molecule is connected to the electrodes by replacing two H atoms of the pyrrole moiety 
and two H atoms on the middle of the armchair facet of the GNR with direct C-C bonds, 
creating a 
side-sharing hexagon-pentagon interface. 

\begin{figure*}[ht]
\includegraphics[width=\textwidth]{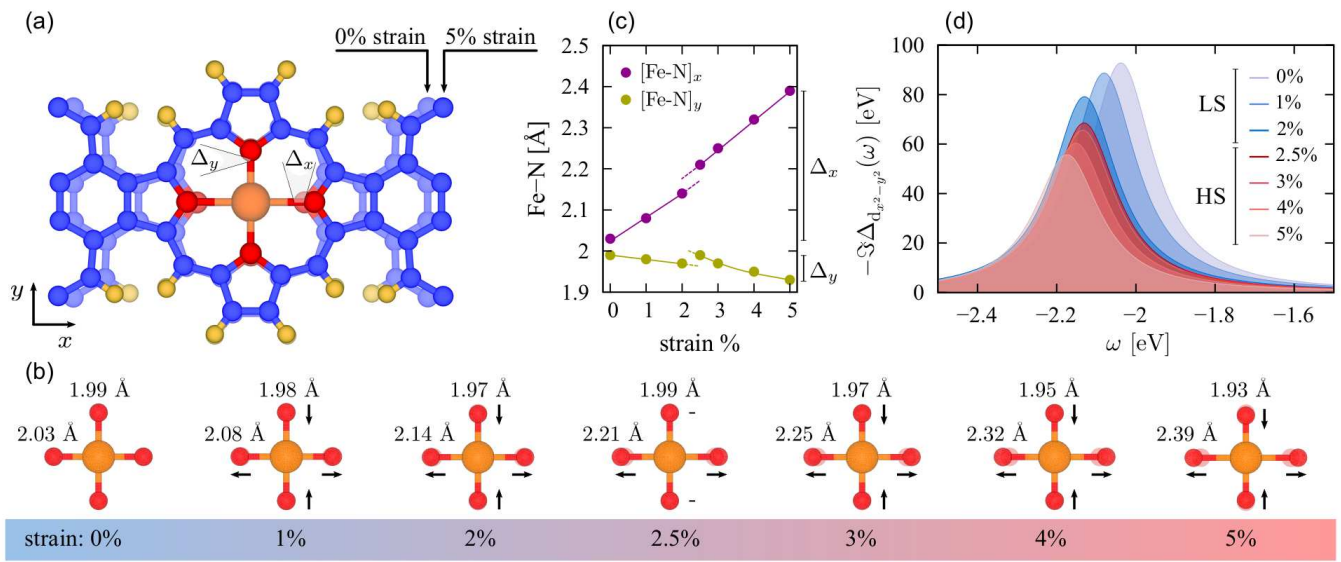} 
\caption{\label{fig:hyb} Effects of strain on the structure of the device and the Fe hybridization.
(a) Atomic positions for 0\% (light) and 5\% (dark) strain. 
The changes in Fe-N bond lengths along the in-plane directions, $\Delta_x=0.36$~\AA \ and $\Delta_y=-0.06$~\AA, are highlighted. 
(b) Asymmetric expansion of the FeN$_4$ core; arrows denote the changes with respect to the 0\% strain bond-length. 
(c) Fe-N bonds along the $x$- and $y$-direction.
(d) Imaginary part of the dynamical hybridization function $\Delta(\omega)$ of the Fe-\dxq\ orbital 
for different strains, with a broadening $\delta = 0.1$~eV. 
}
\end{figure*}

We exploit the magneto-structural relationship 
of the encapsulated FeP molecule~\cite{Scheidt1981,sumprb,sumanta_prl,sumanta_surf} 
as the basis for the device functionality. 
FeP has been demonstrated to realize a spin-crossover (SCO) under various conditions, 
e.g., through surface adsorption,~\cite{fephsau,sumanta_surf,fephsau2,Bhandary2013} 
in molecular assembly,~\cite{Rolf2018} or 
in single-molecule junctions by approaching an STM tip~\cite{KF2018,Kuang2017}, 
all of which create  internal strains in the embedded molecule. 
However, depending on the surface topology and surface-molecule interactions, outer ligand groups, 
metal atom coordination and oxidation state, 
and the nature of external stimuli, the type of magnetic transition, 
\ie low-spin (LS) to high-spin (HS) or {\it vice versa}, 
as well as the amount of required strain, 
vary substantially, often diminishing the 
propensity for a
SCO.~\cite{Li2019,Herper2013} 

In fact, the spin-state of the central Fe atom is the result of a delicate interplay of structural distortions 
in the porphyrin molecule and the Coulomb interaction. 
We manipulate this balance by mechanical strain, as indicated in Fig.~\ref{fig:unit_cell}, 
with the objective of inducing a spin-crossover. 
To effectively include the feedback of a SCO on the atomic structure, we relax the device 
by taking into account the Coulomb interaction within a {\it static} mean-field approximation (DFT+U). With the obtained structures 
we then go beyond this 
approximation and  include 
dynamical
many-body effects within the Fe-$3d$ multiplet in order to obtain a better description of the electronic and transport properties across the SCO. For technical details, see the Supporting Information (SI).

Upon atomic relaxation, the free-standing device undergoes a slight out-of-plane buckling due to an internal stress, caused by the mismatch in C-C bond-length at the interface between the hexagon of the zGNR and the C$_4$N pentagon of the porphyrin ring.
Also, the coordination of the central Fe atom becomes asymmetrical and consequently, the D$_{4h}$ symmetry of the FeN$_4$ core (present in a free molecule) is lifted, as the Fe-N bonds along (perpendicular to) the zGNR coordination direction are elongated (shortened) to [Fe-N]$_x = 2.03$~\AA \ ([Fe-N]$_y = 1.99$~\AA). 

The uniaxial strain applied to the device is quantified by the relative change 
in the unit-cell lattice constant along the $x$-direction. 
The covalent nature of the molecule-zGNR  bonds and the high stretchability of zGNRs~\cite{stretch} 
allow a sizeable expansion of the device. 
A 1\% strain is sufficient to release the internal stress caused by the bonding between FeP and the zGNR electrodes and restore planarity, which is retained for higher strains. 
In Fig.~\ref{fig:hyb}(a) we show how the molecule-zGNR contacts and the molecule itself change due to the applied strain, by comparing structures without and with (5\%) strain. 
Strain stretches the C-C bonds at the contacts as well as the longitudinal Fe-N bonds of FeP, which are identified as the softest in the system, by far. As expected, the changes in the Fe-N bond-length is direction-dependent and the [Fe-N]$_x$ distances expand while the [Fe-N]$_y$ distances shrink, see Fig.~\ref{fig:hyb}(b,c), further enhancing the asymmetry of the active molecule. 
Between 2\% and 2.5\% strain, the evolution of interatomic distances is discontinuous as the FeN$_4$ core suddenly expands in both $x$- and $y$-direction. 
This phenomenon is, as we shall see, the structural signature of the system switching from a LS ($S=1$) to a HS ($S=2$) ground-state.

We now take the perspective of the central Fe-ion, as it carries the magnetic degrees
of freedom responsible for the device functionality. 
More specifically, we quantify how the above structural changes modify the Fe-ligand coupling, 
by considering the dynamical hybridization function~\cite{bullaRMP80,schuelerEPJBST226,gandusJCP153} 
\begin{equation}
    \Delta_{\ell}(\omega) = \sum_n \frac{|V_{\ell n}|^2}{\omega-\epsilon^b_n+\imath\delta}
    \label{eq:hyb}
\end{equation}
where $\ell=\{\dxy, \dyz, \dzq, \dxz, \dxq\}$ denote the Fe-$3d$ orbitals, 
$\epsilon^b_n$ the energies of the molecular orbitals 
of the ligands, which include both the porphyrin ring and the zGNR leads, 
and $V_{\ell n}$ their coupling. 
The component $\Delta_{\ell}(\omega)$ with $\ell=$ \dxq\ is the strongest 
due to the large axial overlap between \dxq\ and N-2$p$ orbitals.
The second strongest contribution comes from the \dyz\ orbital, 
whereas the \dxz\ one is suppressed by the elongation of the [Fe-N]$_x$ distance and 
is already insignificant for the description of the SCO (see Fig.~S1 in the SI). 
In the following we focus on the \dxq\ component of $\Delta(\omega)$, which is displayed in Fig.~\ref{fig:hyb}(d). At 0\% strain, $\Im\Delta_{\dxq}$ is characterized by a broad resonance around $\omega \sim -2.0$~eV. 
As a function of strain, it shifts towards higher binding energy and its height decreases, 
a hallmark of the weakening metal-ligand coupling. 
A sudden drop between 2\% and 2.5\% strain reflects again the structural discontinuity connected with the SCO.

\begin{figure}[!t]
\includegraphics[width=0.5\textwidth]{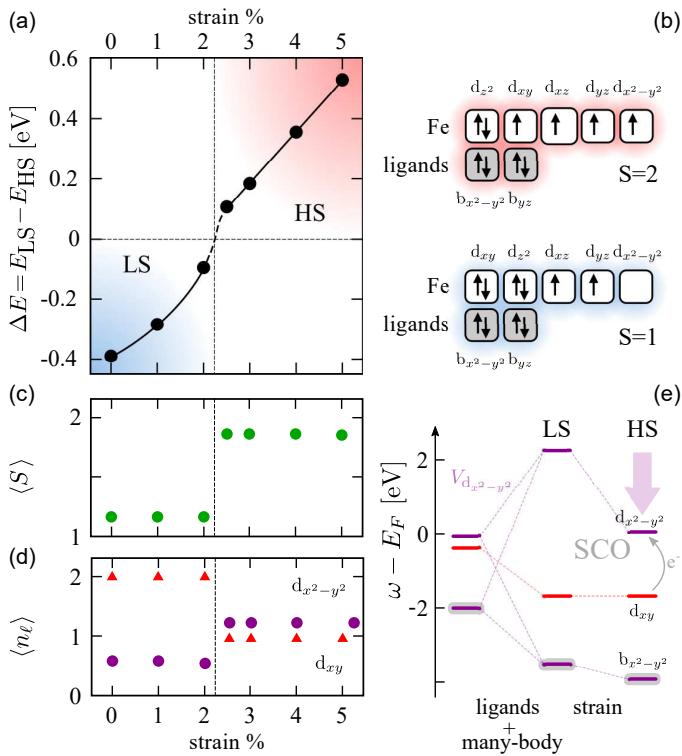}
\caption{\label{fig:LSHS} Spin-state switching of FeP in a device. (a) Spin-transition energy. 
(b) Many-body ground state configurations projected onto the 
Fe-$3d$ and ligand orbitals.  
(c) Average total spin $\left\langle S\right\rangle$, and (d) average occupations of the \dxq\ and \dyz\ orbitals, as a function of strain. 
(e) Molecular level-diagrams showing the mechanism of the strain-induced spin-transition. }
\end{figure}

%
To reveal the connection between structural and magnetic changes under strain, we now 
take into account many-body effects 
in the Fe-$3d$ multiplet embedded into its surrounding via the hybridization function, \eref{eq:hyb}. 
The latter is restricted to the dominant \dxq\ and \dyz\ channels which are treated as bath degrees of freedom. 
We then solve the ensuing realistic Anderson impurity model that incorporates the full Coulomb tensor~\cite{valliPRR2} 
with an exact diagonalization (ED) technique (for details, see the SI). 

In Fig.~\ref{fig:LSHS} we show the evolution of the electronic properties under strain. 
In Fig.~\ref{fig:LSHS}(a), the spin-transition energy $\Delta E =E_{LS}-E_{HS}$, 
i.e., the energy difference between the lowest-lying many-body states in the LS and HS sectors,  
displays a discontinuity as a function of mechanical strain.
We identify the system to undergo a transition from a LS to a HS state (sign change in $\Delta E$) 
at the same critical strain (2-2.5\%) 
for which DFT+U accounted for the structural transition. 
The corresponding electronic configurations are schematically shown in Fig.~\ref{fig:LSHS}(b) 
in the basis of the Fe-$3d$ and the two ligand orbitals. 
The LS and HS states differ for the occupation of the \dxy\ and \dxq\ orbitals, 
and the alignment of spins on different orbitals follows the Hund's rule.  
The above scenario is confirmed by the discontinuous evolution with strain 
of the total spin $\langle S \rangle$ projected onto the Fe-$3d$, 
between $\langle S \rangle_{\textrm{LS}} \sim 1.2$ (strain $\leq 2$\%) and $\langle S \rangle_{\textrm{HS}} \sim 1.9$ (strain $>2$\%).
Concomitantly, the orbital occupation $\langle n_{\ell} \rangle$ 
shows a charge transfer from the \dxy\ to the \dxq\ orbital. 
The slight deviation from a pure $\langle S \rangle_{\textrm{LS}} = 1$ 
and $\langle S \rangle_{\textrm{HS}} = 2 $ can be mainly attributed 
to the projection on to the Fe-3$d$ orbitals, as explained 
previously.~\cite{sbnip} 
The expectation values of the observables are obtained as a quantum mechanical and thermal average 
from many-body calculations at $T=300$~K, 
indicating that the SCO survives up to room temperature. 
The mechanism underlying the SCO is represented schematically by the level diagram in Fig.~\ref{fig:LSHS}(e) 
and can be understood in terms of the competition between the Coulomb interaction and the ligand  field. 
Both in the gas phase and in the device at 0\% strain, 
the anti-bonding orbital with predominant \dxq\ character 
lies far above the Fermi level, unoccupied, due to the strong hybridization to the ligands. 
As the Fe-N bonds soften along the strain direction, the splitting between bonding and anti-bonding orbitals 
is strongly reduced. At a critical strain, it becomes energetically favorable 
to promote an electron from the doubly-occupied \dxy\ to the anti-bonding \dxq\ orbital, 
in order to evade the intra-orbital Coulomb repulsion. 
At the same time, the Hund's exchange couplings within the Fe-$3d$ multiplet suppress 
inter-orbital ($\ell\ne \ell'$) double occupations with opposite spins 
$\langle n_{\ell\uparrow} n_{\ell'\downarrow}\rangle \ll \langle n_{\ell\sigma} n_{\ell'\sigma}\rangle $ 
thus stabilizing a HS configuration.~\cite{valliPRR2}

\begin{figure*}[!t]
\includegraphics[width=\textwidth]{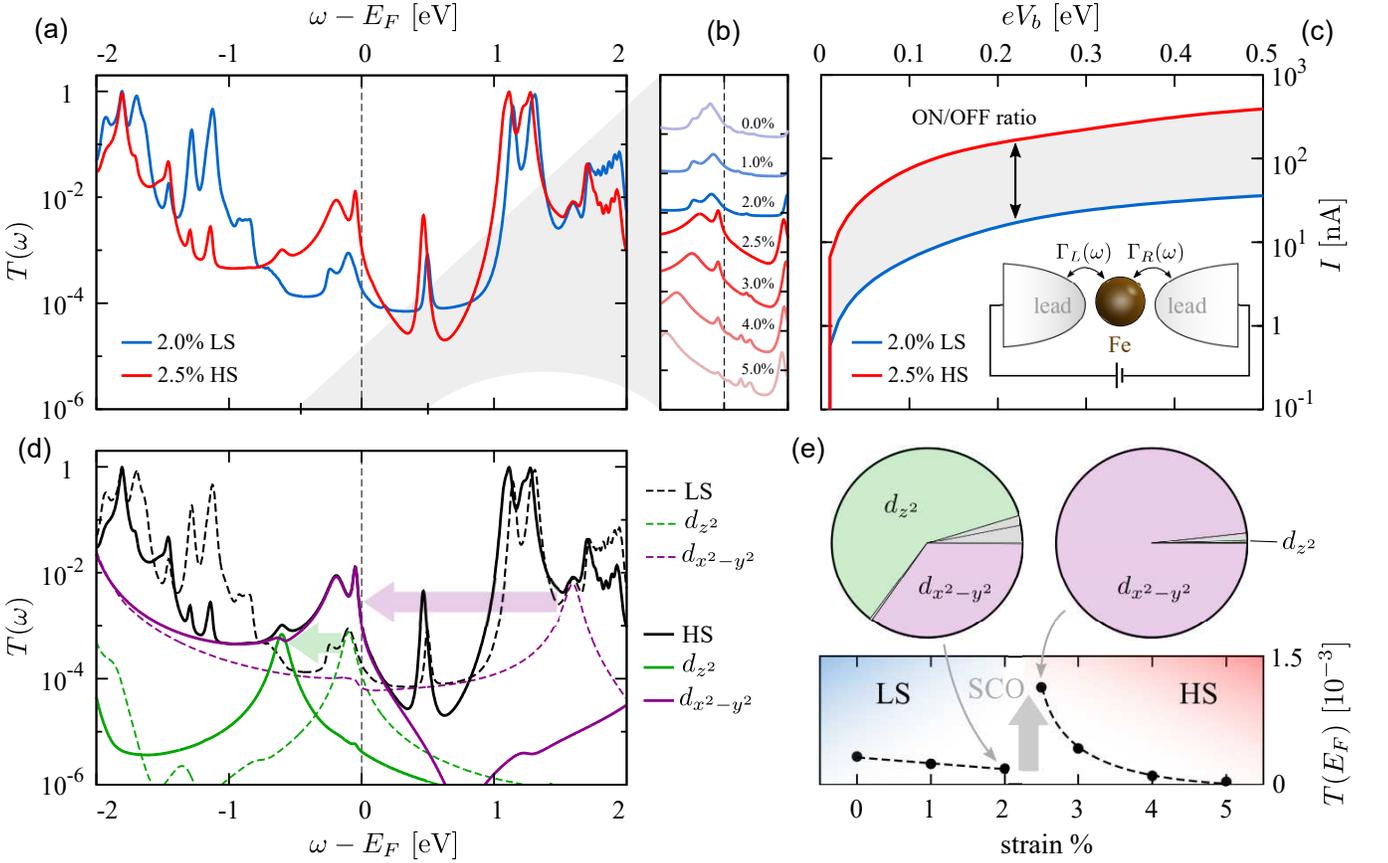}
\caption{\label{fig:Te_LSHS} (a) Transmission function in the LS and HS state and 
(b) evolution of the transmission close to $E_F$ across the SCO. 
(c) I-V characteristics in the LS and HS state, demonstrating the ON/OFF ratio $I_{HS}/I_{LS}>10$. 
(d) Transmission function (black lines) and its orbital-resolved contributions 
from \dxq\ and \dzq\ orbitals (color lines), which dominate the charge transport close to $E_F$. 
The arrows highlight the redistribution of electronic spectral weight associated to the SCO. 
(e) Evolution of the transmission at the Fermi level as a function of strain. 
The pie charts highlight the \dxq\ and \dzq\ orbital character in the LS and HS state (color) 
while all other orbitals are displayed in gray. 
}
\end{figure*}
%

One can estimate the Coulomb contribution to the spin-transition energy 
between the electronic configurations in Fig.~\ref{fig:LSHS}(b). 
Within a density-density approximation of the Coulomb tensor (see SI for an extended discussion) 
$\Delta_{\textrm{Coulomb}} = 2J_1 + 3J_3$. 
The system gains energy in the HS state associated with two exchange interaction parameters: 
$J_1$, that couples both the \dxy\ and \dxq\ to the \dxz\ and \dyz\ orbitals, and
$J_3$, that couples the \dxy\ and \dxq\ orbitals. 
We find $\Delta_{\textrm{Coulomb}} \approx 2.83$~eV to be of the same order of magnitude 
as the ligand field between the \dxz\ and \dxq\ orbitals. 
The competition between these two energy scales, which in the full calculation 
is represented by $\Delta E$ in Fig.~\ref{fig:LSHS}(a), is the root cause of the SCO in FeP.

%
The change of the spin state across the SCO has practical implications: It drastically changes the charge transport properties through the device. 
In order to demonstrate that, we map the device onto an effective quantum-dot model.  
We calculate the Landauer transmission function
\begin{equation} \label{eq:landauer}
 T(\omega) = Tr \big[ \Gamma^L G^{\dagger} \Gamma^R G^{\phantom{\dagger}}\big], 
\end{equation} 
that describes the probability of an electron to be transferred coherently through the Fe atom.
Here $G$ denotes the Green's function of the Fe-$3d$ multiplet, which includes the many-body effects 
stemming from the Coulomb interactions~\cite{meirPRL68,jacob15,droghettiPRB95,valliNL18,valliPRB100} 
and $\Gamma^{\alpha} = \imath(\Sigma^{\alpha}-\Sigma^{\dagger\alpha})$,  
describes the coupling to the rest of the device, 
when the hybridization function $\Delta(\omega) = \Sigma^L(\omega)+\Sigma^R(\omega)$ 
is separated into components to the left (L) and the right (R) of the Fe atom (see the SI for details). 

The simulated transmission function $T(\omega)$ in the LS- and HS-state is shown in Fig.~\ref{fig:Te_LSHS}(a) 
for strains across the SCO, while (b) shows its behavior at low-energy for all strain values.
At the Fermi level we identify a significant 
enhancement 
of the conductance from about $2 \times 10^{-4} G_0$ at $2\%$ strain 
to values $> 10^{-3} G_0$ at $2.5\%$ strain, 
where $G_0=e^2/h$ is the conductance quantum. 
This jump is mirrored by an increase in the electric current (per spin) evaluated as 
\begin{equation}\label{eq:IV}
 I = \frac{e}{h} \int_{-\infty}^{\infty} d\omega \ T(\omega) \big[ f_L(\omega) - f_R(\omega) \big],
\end{equation}  
where $e$ is the electron charge, $h$ is the Planck constant, and
$f_{L/R}(\omega)=f(\omega-\mu_{L/R})$ is the Fermi-Dirac distribution function of the leads
symmetrically displaced by the bias voltage $\mu_L-\mu_R=V_b$. 
The junction displays a linear transport characteristics and a large ON/OFF ratio $I_{HS}/I_{LS}>10$, 
which is remarkably independent on the external bias voltage, 
as shown in Fig.~\ref{fig:Te_LSHS}(c). 
An orbital dependent analysis of the transmission function, reported in Fig.~\ref{fig:Te_LSHS}(d,e)
allows us to unambiguously trace back the changes in the transport properties to the SCO. 
In the LS state, the orbital character of the transmission function is prevalently $\dzq$ 
due to the hybridization of the Fe with the N atoms along the $y$ direction (perpendicular to the transport),
with  about 30\% admixture of \dxq \ character.
In the HS state, the nonbonding \dzq \ orbital shifts to lower energy 
and the transmission function is instead dominated by a narrow resonance 
with \dxq \ orbital character just below the Fermi level, 
which appears as a consequence of the partial occupation of the corresponding Fe orbital. 
As a result, the transmission function at the Fermi level displays 
a sharp increase as a function of strain at the SCO. 
Hence, the transport properties allow to discriminate between the two spin states. 
Similar behavior is observed for switches based on photo-induced molecular isomerization, 
where the change in the transport properties are ascribed to changes in the chemical configuration.~\cite{zhangCSR44}


In summary, we have demonstrated the potential of a molecular spin-switch device, where a mechanically driven spin-crossover results in a bias-independent enhancement of the current through the device by one order of magnitude. 
The key advantage of the proposed design is the organic anchoring of the active FeP molecule. The strong covalent bonding to the zGNR electrodes assures a high reliability and reproducibility. The relatively smaller contact-resistance 
is also expected to enable the proposed device to identify small conductance differences in electrical measurements. 
Recent technical breakthroughs, 
such as dash-line lithography~\cite{cao12} or on-surface-synthesis~\cite{he17,li18} 
make it 
possible to anchor molecules with graphene-nano structures with high atomic-precision. 
These graphene-molecule-graphene 
junctions have already shown remarkable functionalities,
including magnetism,~\cite{jia15,li20,li18,Li2019} 
and can be engineered to work in reversible mechanical break-junction setups.~\cite{Caneva2018} 
These early successes strongly suggest that our proposed device, which  
couples magnetic and structural degrees of freedom
in a fully planar geometry, 
is experimentally realizable, adding a new building block for organic spintronic devices.

\vspace{3mm}

We acknowledge financial support from the Science Foundation Ireland [19/EPSRC/3605] and the Engineering and Physical Sciences Research Council [EP/S030263/1], European Research Council (Consolidator Grant No. 617196 CORRELMAT), Austrian Science Fund (FWF) through project `LinReTraCe’ P 30213-N36 (JMT, AV) and project P 31631 (AV).

\vspace{3mm}

\paragraph*{Supporting Information.} 
Details for {\it ab-initio} and many-body calculations; 
embedding of the Fe-$3d$ multiplet; 
benchmark against QMC impurity solver; 
transport calculations (PDF).

\bibliographystyle{apsrev}
\bibliography{arXiv}

\end{document}


\onecolumngrid
\thispagestyle{empty}

\setcounter{equation}{0}
\setcounter{figure}{0}
\setcounter{table}{0}
\setcounter{page}{1}
\makeatletter
\renewcommand{\thefigure}{S\arabic{figure}}
\renewcommand{\thepage}{S-\arabic{page}}

\newcommand*\mycommand[1]{\texttt{\emph{#1}}}

\begin{center}
  \textbf{\large Supporting Information: \\Designing a mechanically driven spin-crossover molecular switch \\via organic embedding}\\[.2cm]
  S.~Bhandary,$^{1,*}$ J.~M.~Tomczak,$^{2}$ A.~Valli$^3$\\[.1cm]
  {\itshape ${}^1$School of Physics, Trinity College Dublin, The University of Dublin, Dublin 2, Ireland}\\
  {\itshape ${}^2$Institute of Solid State Physics, Vienna University of Technology, 1040 Vienna, Austria}\\
  {\itshape ${}^3$Institute for Theoretical Physics, Vienna University of Technology, 1040 Vienna, Austria}\\
\end{center}

\subsection*{DFT computational details} 
We have used Vienna ab-initio Simulation Package~\cite{SI_vasp} (VASP) for the DFT calculations. The device is periodic in the x-direction. In the  y-direction the separation from the periodic image is more than 16 \AA. The vacuum in the z-direction is 20 \AA. Ionic positions of all the structures have been optimized. 
The plane-wave projector augmented wave basis was used in the Perdew-Burke-Ernzerhof generalized gradient approximation (PBE-GGA) for the exchange correlation potential. The plane-wave cutoff energy used was 400~eV. A 10$\times$1$\times$1 Monkhorst Pack k-mesh was used for the integration in the Brillouin zone. Atoms were relaxed until the Hellmann-Feynman forces were  below 0.01~eV/\AA. 
As electronic correlations play a crucial role in determining the SCO, 
it is reasonable to include their feedback also in the structural relaxation. 
We have  performed 
structural relaxation within the DFT+U formalism~\cite{SI_licldau} with interaction parameters U=4.0~eV and J=1.0~eV, 
which take into account the static contributions of the Coulomb repulsion, 
and, to some extent, can already describe the SCO.
It has to be noted that the parameters for the many-body calculations, i.e., the hybridization functions and crystal fields (see also below), are obtained from non-spin polarised (i.e., {\it sans} U) calculations using those relaxed structures (obtained in DFT+U) while dynamical correlation effects are incorporated within the explicit many-body techniques.

\subsection*{Many-body effects: DFT++}
To investigate the molecular spin-crossover, we have employed a combination of DFT 
and many-body techniques within a multi-orbital Anderson impurity model (AIM), 
which is usually referred to as a DFT++ method.~\cite{SI_dft++}

\subsubsection*{Anderson impurity model}
A realistic description of the complete system is obtained within DFT. 
This allows to extract \emph{ab-initio} parameters 
to describe a correlated sub-space, i.e., the impurity, corresponding to the Fe-$3d$ multiplet, 
which is coupled to a bath, corresponding to the rest of the system, 
including both the porphyrin ligands and the zGNR leads, via a retarded hybridization function. 
The impurity is supplemented with a Coulomb interaction, and the resulting many-body problem 
can be solved numerically. The Hamiltonian of the AIM can be expressed as 
\begin{equation}\label{eq:ham}
 \begin{split}
 H =              \sum_{ij}   \epsilon^{d}_{ij} 
                            d^{\dagger}_{i\sigma} d^{\phantom{\dagger}}_{j\sigma}
   + \frac{1}{2} &\sum_{ijkl} \sum_{\sigma\sigma'} U_{ijkl} 
                            d^{\dagger}_{i\sigma} d^{\dagger}_{j\sigma'} 
                            d^{\phantom{\dagger}}_{l\sigma'} d^{\phantom{\dagger}}_{k\sigma} \\
   +             &\sum_{im} \sum_{\sigma} (V_{im} c^{\dagger}_{m\sigma} d^{\phantom{\dagger}}_{i\sigma}
                                         + h.c.) 
   +              \sum_{m} \sum_{\sigma} \epsilon_m 
                                       c^{\dagger}_{m\sigma} c^{\phantom{\dagger}}_{m\sigma}, 
 \end{split}
\end{equation}
where $d^{\phantom{\dagger}}_{i\sigma}$ ($d^{\dagger}_{i\sigma}$) denote 
the annihilation (creation) operators of an electron 
in impurity (Fe-$3d$) orbital $i$ with spin $\sigma$, 
and $c^{\phantom{\dagger}}_{m\sigma} $($c^{\dagger}_{m\sigma}$) denote 
the annihilation (creation) operators of an electron 
in bath orbital $m$ with spin $\sigma$ and energy $\epsilon_m$. 
The coupling between Fe-$3d$ and bath orbitals is given by $V_{im}$. 
The matrix $\epsilon^{d}_{ij}$ describes the 
crystal field and $U_{ijkl}$ represents the full Coulomb tensor 
within the Fe-$3d$ multiplet.  
The rotationally-invariant Coulomb interaction is parametrized 
via the Slater radial integrals~\cite{SI_slater1929,SI_slater1960} 
$F^0$, $F^2$, and $F^4$, 
such that $U=F^0$ and $J=\frac{1}{14}(F^2+F^4)$, with the ratio $F^4/F^2=0.625$, 
yielding a spherically symmetric tensor.~\cite{SI_karolakPhD,SI_valliPRR2} 
In the presence of a crystal field, the spherical symmetry is lifted. 
This effect could be taken into account,  e.g., within the constrained random phase approximation,~\cite{SI_jacob15}  
but we do not expect it to change any of the conclusions of our analysis.

\subsubsection{Embedding the Fe-atom: Hybridization function}
\begin{figure}[t]
\includegraphics[width=1.0\textwidth]{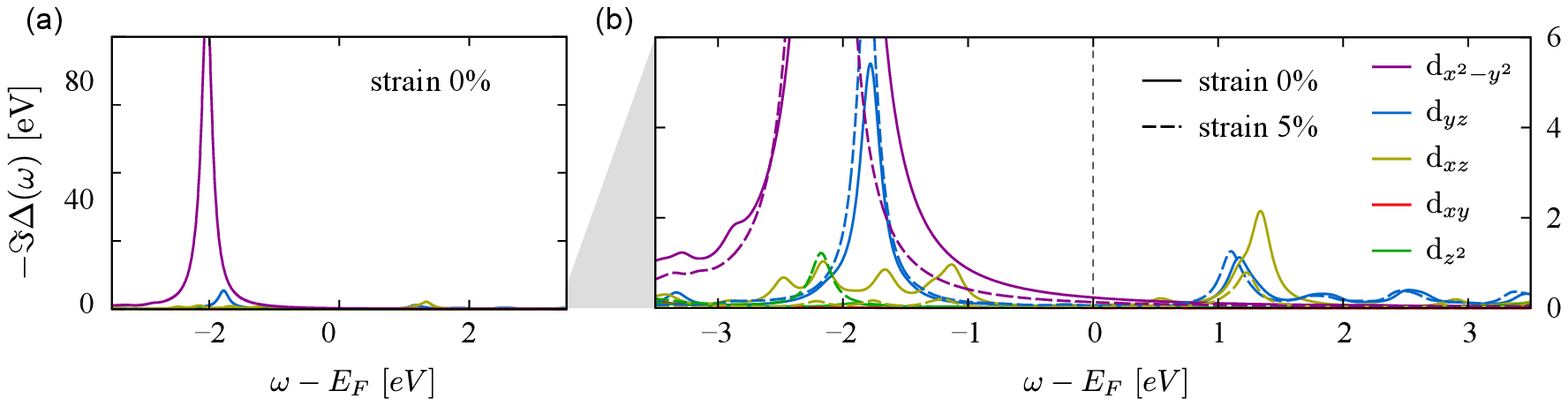}
\caption{\label{fig:hyb_full} (a)  Orbital-resolved hybridization function $\Delta(\omega)$ for the unstrained device. 
The \dxq\ peak at $\epsilon^b=-2.03$~eV dominates the frequency structure. 
(b) Evolution of the hybridization function across the SCO, at 0\% (solid) and 5\% (dashed lines) strain. }
\end{figure}

In order to calculate the hybridization function \emph{ab-initio}, the Kohn-Sham Green's function G$_{KS}$ 
is calculated from the Lehmann representation 
\begin{equation}
G_{KS}(\omega)=\sum_{nk}\frac{\ket{\psi_{nk}}\bra{\psi_{nk}}}{\omega+i\delta-\epsilon_{nk}},
\end{equation}
where $\psi_{nk}$ and $\epsilon_{nk}$ are the Kohn-Sham eigenstates and eigenvalues 
for band $n$ and reciprocal-space point $k$, 
while $\delta$ is an infinitesimal broadening, indicating $G_{KS}$ to be the retarded propagator.
This Green's function is then projected onto a local (impurity) propagator $G_{0}$, 
evaluated on atom-centered, localized orbitals $\chi_{i}$. 
In this basis, the local Green's function reads 
\begin{equation}
             G^{ij}_{0}(\omega)=\sum_{nk}\frac{\tilde{P}^{i}_{nk}{(\tilde{P}^{j}_{nk}})^{*}}{\omega+i\delta-\epsilon_{nk}}
\end{equation}
where $P^{i}_{nk}=\braket{\chi_{i} | \psi_{nk}}$ 
are projection matrices normalized as 
\begin{align}
\tilde{P}^{i}_{nk}=\sum_{j}[O(k)]^{-1/2}P^{j}_{nk},
\end{align}
with the overlap operator
\begin{align}
O_{ij}(k)=\sum_nP^{i}_{nk}(P^{j}_{nk})^*.
\end{align}
Finally, the hybridization function is calculated from the local impurity Green's function from the expression
 \begin{equation}
            \Delta_{ij}(\omega)=\big[\omega+i\delta\big]\delta_{ij}-\epsilon^d_{ij}-\big[G^{-1}_{0}\big]_{ij}(\omega).
\end{equation}
Note that the hybridization function includes also off-diagonal terms 
if the projected Fe-$3d$ orbital are not orthogonal (i.e., $\epsilon^d_{i \neq j} \neq 0$). 
In the current cases $\Delta_{i\neq j} \ll \Delta_{ii}$ for any pair ($i$, $j$), 
and neglecting these contributions yields an expression equivalent to the one given in the manuscript, 
yet projected on the localized Fe-$3d$ orbitals. 
The diagonal elements of the hybridization function $\Delta_{ii}(\omega)$ are shown in Fig.~\ref{fig:hyb_full}, 
clearly indicating that the \dxq\ is the dominant contribution among them.

\subsubsection{Numerical solution: ED vs. QMC}
\begin{figure}[t]
\includegraphics[width=0.8\textwidth]{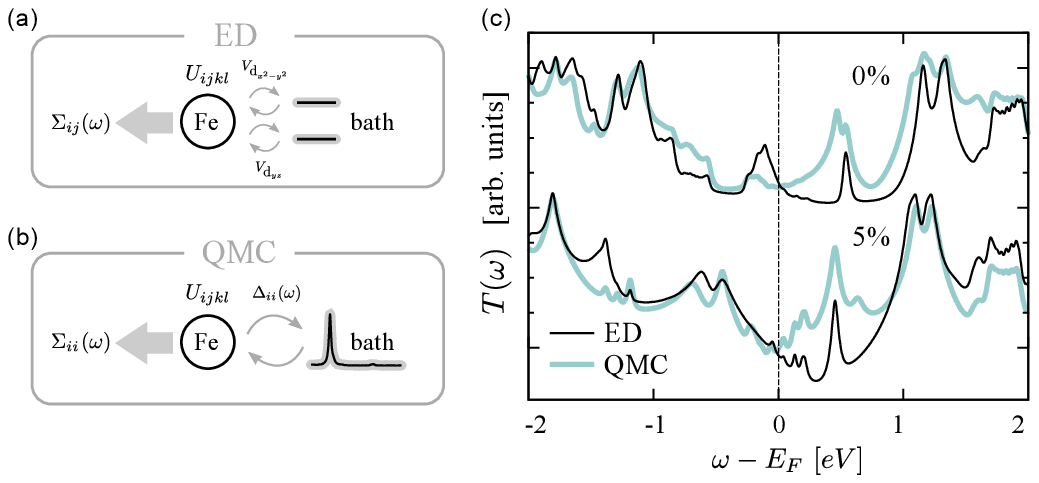}
\caption{\label{fig:Te_EDvsQMC} Schematic representation of (a) ED and (b) QMC impurity solvers. 
(c) Transmission function at 0\% and 5\% tensile strain, obtained from many-body calculations 
with both solvers. The curves are shifted vertically for a better visibility. }
\end{figure}

The AIM in Eq.~(\ref{eq:ham}) can be solved with several numerical techniques, 
each with their advantages and disadvantages. 
Here we adopt exact diagonalization (ED) for the results presented in the manuscript 
and continuous-time Quantum Monte Carlo~\cite{SI_gullRMP83} (QMC) as a benchmark. 
A schematic representation of the ED and QMC algorithms is shown in Figs.~\ref{fig:Te_EDvsQMC}(a,b) 
while results for the corresponding transmission function, 
are shown in Fig.~\ref{fig:Te_EDvsQMC}(c). 
This comparison highlights a remarkable agreement for both 0\% and 5\% tensile strain, 
which correspond to LS and HS state configurations, respectively. 
This analysis shows that the physical scenario underlying the SCO in the zGNR-FeP-zGNR device 
is robust with respect to the details of our numerical calculations. \\

\noindent
{\bf ED solver.} As the ED solver is based on the Hamiltonian formalism, 
the dynamical hybridization function is discretized, 
i.e., approximated by a finite number of bath sites. 
The hybridization function in Fig.~\ref{fig:hyb_full} is dominated by the diagonal \dxq\ contribution, 
and the second strongest hybridization is the \dyz\ one, 
while off-diagonal contributions are negligible. 
Hence, for the ED calculations, it is sufficient to fit those two contributions with one pole each. 
This procedure yields the bath orbital energies $\epsilon^b$ and the corresponding coupling $V$ 
for the \dxq\ and \dyz\ orbitals listed in Tab.~\ref{tab:APs}. 

\begin{table}[b!]
\centering
 \caption{Anderson parameters of the ED calculations, for different values of uniaxial strain. }
 \label{tab:APs}
\renewcommand{\arraystretch}{1.2}
\begin{tabular}{ c ccccccc } 
 \hline 
 \multirow{2}{*}{Parameters [eV] \ } & \multicolumn{7}{c}{Strain [\%]} \\ 
 \cline{2-8} & 0  &  1  &  2  &  2.5  &  3 &  4 &  5 \\ 
 \hline
 \hline
 $V_{\dxq}$              &\ 3.04 &\ 2.97 &\ 2.80 &\ 2.61  &\ 2.56  &\ 2.45  &\ 2.35  \\ 
 $\epsilon^b_{\dxq}$     & -2.03 & -2.08 & -2.13 & -2.13  & -2.14  & -2.15  & -2.18  \\ 
 \hline
 $V_{\dyz}$              &\ 0.74 &\ 0.76 &\ 0.79  &\ 0.76  &\ 0.79  &\ 0.83  &\ 0.89  \\ 
 $\epsilon^b_{\dyz}$     & -1.78 & -1.80 & -1.81  & -1.80  & -1.79  & -1.79  & -1.81  \\ 
 \hline
 \hline
\end{tabular}
\end{table}

In the solution of the many-body problem, we include a fully localised limit (FLL) double counting correction.~\cite{SI_Lichtenstein_LDA+U,SI_Sawatzky_DC}, according to 
\begin{equation}
 \epsilon^{\textrm{DC}} = U\bigg(n-\frac{1}{2}\bigg) - \frac{J}{2} \bigg(n-1\bigg)
\end{equation}
with $n$ being the total impurity occupation. 
However, instead of considering the DFT-occupations, we obtain $\epsilon^{\textrm{DC}}$ in a charge self-consistent manner, i.e., we start with an initial guess of $\epsilon^{\textrm{DC}}$, obtain new occupations in ED and calculate a new $\epsilon^{\textrm{DC}}$. The process is iterated until we obtain a convergence over the total occupation.  
The solution of the AIM yields a retarded self-energy matrix $\Sigma_{ij}(\omega)$
that takes into account the many-body effects within the Fe-$3d$ multiplet. 
The corresponding many-body Green's function is given by
\begin{equation} \label{eq:gf}
 G^{-1}_{ij}(\omega) = \big[ \omega + \imath\delta + \epsilon^{\textrm{DC}}\big]\delta_{ij} - \epsilon^d_{ij} 
                     - \Delta_{ij}(\omega) 
                     - \Sigma_{ij}(\omega),
\end{equation}
which in turn allows to evaluate the transmission function 
and investigate the transport properties of the device, see below. 

Note that in the DFT++ framework it is possible to obtain a paramagnetic solution, 
so that the Hamiltonian, the hybridization function, the self-energy, 
and eventually also the many-body Green's function, are not spin-polarized. 
In contrast to, e.g., DFT+U, this scheme allows to describe a fluctuating local moment on the Fe atom
and the transport properties across the SCO in the absence of any static magnetic order. \\

\noindent
{\bf QMC solver.} Within the QMC algorithm, the dynamical nature of the hybridization function can be fully taken into account, 
and the double counting correction is implemented by fixing the occupation of the Fe-$3d$ manifold 
$n = \sum_{i\sigma}\langle n_{i\sigma} \rangle$ to the values 
$n^{\textrm{LS}}=6.7$ and $n^{\textrm{HS}}=6.4$~electrons. These values are chosen in correspondence with the impurity occupations in ED for the two different spin-states. 
Due to technical complications, it is convenient to neglect the off-diagonal elements 
of the Hamiltonian and the hybridization function. 
As a result, the QMC self-energy becomes diagonal in orbital space, 
i.e., $\Sigma_{ij}(\omega)=\delta_{ij} \Sigma_{ii}(\omega)$. 
The remarkable agreement found between the QMC and ED data allows us to verify {\it a posteriori} that 
(i) the off-diagonal components of the self-energy have a negligible effect on the result, and 
(ii) the hybridization function is well approximated with two ED bath sites. 
In the context of a free FeP molecule, the convergence of spin-transition energy with the number of bath sites was checked in a previous publication,~\cite{SI_sumprb} further justifying our choice here.
Let us also note that, in the case of QMC, an additional step is required, namely to analytically continue 
the Matsubara self-energy to the real-frequency axis via a maximum entropy method.~\cite{SI_jarrell1996} 

\subsubsection*{Competition between ligand crystal field and Coulomb interaction}
In order to have a better grasp of the competition between the ligand crystal field 
and the Coulomb interaction, one can extract an effective energy level diagram from the ED calculations. 
We consider an effective 7$\times$7 Hamiltonian for the five Fe-$3d$ and the two ligand orbitals, 
which in block-matrix form can be represented as
\begin{equation}\label{eq:eff-ham}
 H_{\textrm{eff}} = 
 \begin{pmatrix}
  \epsilon^d & V \\
  V^\dagger        & \epsilon^b
 \end{pmatrix}
 +
 \begin{pmatrix}
  \Sigma^{\infty} & 0 \\
  0                  & 0 
 \end{pmatrix}.
\end{equation}
In this notation, $\epsilon^d$ is a 5$\times$5 matrix representing the crystal fields of the Fe-$3d$ multiplet, 
$\epsilon^b$ is a 2$\times$2 (diagonal) matrix with the ligand orbital energies,  
and $V$ is a 2$\times$5 matrix with only two non-zero elements, corresponding to the couplings 
$V_{\dxq}$ and $V_{\dyz}$ defined above, 
while $\Sigma^{\infty}=\Re\Sigma(\omega \rightarrow \infty)$ is the static contribution 
to the many-body self-energy in the Fe-$3d$ subspace. 
Since the parameters of the model depend on strain, we can define a family of Hamiltonians. 
For a given strain, the diagonalization of the respective Hamiltonian yields the eigenvalues $\{\epsilon_i\}$ 
corresponding to molecular orbitals (MOs) with Fe-$3d$ and ligand mixed character. 

\begin{figure}[t]
\includegraphics[width=0.8\textwidth]{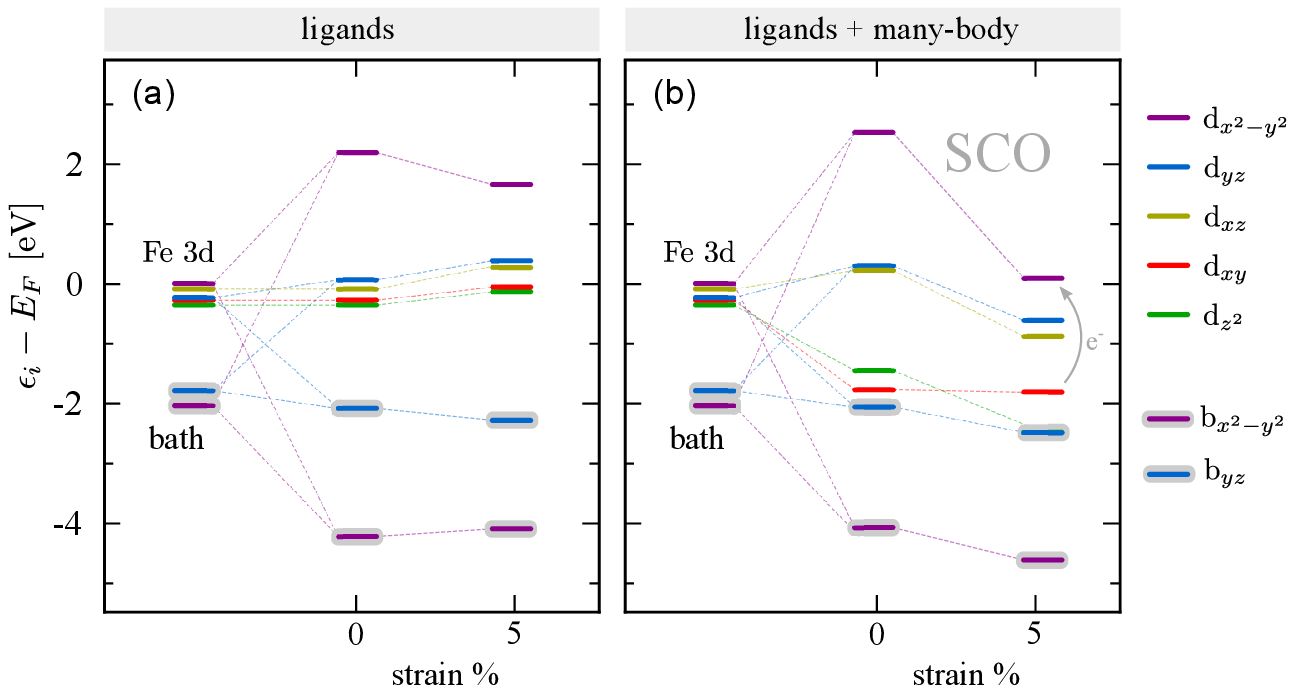}
\caption{\label{fig:level_diagram} Effective level diagram obtained 
within the Fe-$3d$ multiplet with two bath site model. 
Effect of the ligand field alone (a) and together with many-body effects (b).  
The latter is necessary to trigger the SCO at realistic values of tensile strain. }
\end{figure}

In Fig.~\ref{fig:level_diagram}(a) we show the MO eigenvalues obtained by considering the effect of the ligands 
{\it without} the self-energy contribution. 
Starting from the isolated subspaces of the unstrained device, due to the couplings 
between the \dxq\ and \dyz\ orbitals and their bath sites, the system creates bonding and anti-bonding MOs. 
However, tensile strain has surprisingly very little effect on the energy level diagram. 
Even at 5\% strain, the anti-bonding MO with predominant \dxq\ character 
is still far above the Fermi level, $\epsilon_{\dxq} \approx 1.8$~eV. 
This suggests that the structural changes alone are not able to trigger the SCO. 
Including the self-energy contributions the scenario changes substantially, 
as shown in In Fig.~\ref{fig:level_diagram}(b). 
The ligand field splitting $\Delta_{\textrm{LF}}=\epsilon_{\dxq}-\epsilon_{\dxy}$ 
is reduced upon strain 
and at the same time $\epsilon_{\dxq} \approx E_F$, thus triggering the SCO. 

\begin{figure}[b]
\includegraphics[width=0.7\textwidth]{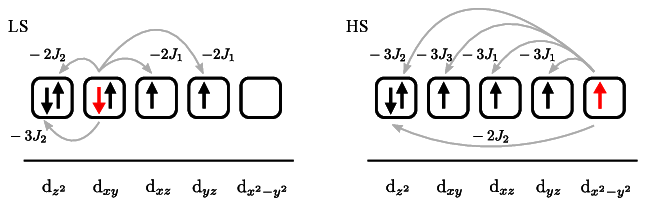}
\caption{\label{fig:dCoulomb}  
Within the density-density approximation, the Coulomb energy difference between the atomic LS and HS states
is given by $\Delta_{\textrm{Coulomb}}=2J_1+3J_3$.  
Since $U$ is the same for all orbitals, the energy difference is obtained from the variation 
of exchange interactions (grey arrows) 
associated to the electron promoted from the \dxy\ to the \dxq\ orbital across the SCO (marked in red). }
\end{figure}

It is interesting to compare the ligand fields to the Coulomb energy difference $\Delta_{\textrm{Coulomb}}$ 
between the atomic LS and HS states. 
Although we consider the full Coulomb tensor $U_{ijkl}$ in the numerical calculations, 
an intuitive picture can already be obtained within a (simplified) density-density approximation, 
under the assumption of a cubic ligand field. 
Then, the Coulomb interaction reduces to 
\begin{equation} \label{eq:Udd}
 \begin{split}
    U_{iiii;\sigma\sigma} & = U_0 \\
    U^{i \neq j}_{ijji;\sigma\sigma'} & = U_0 - 2J_{ij} - J_{ij}\delta_{\sigma\sigma'},
 \end{split}
\end{equation} 
and is given in terms of an intra-orbital interaction $U_0$ (which is the same for all orbitals) 
and four exchanges $J_{ij} = \{J_1, J_2, J_3, J_4 \}$. 
Including the term of the Coulomb tensor beyond the density-density approximation 
cannot be done at this intuitive level, because they conserve neither 
the spin-orbital charge $n_{i\sigma}$ nor the spin projection $S^{z}_{i}$ quantum numbers. 
An extensive discussion of the Coulomb parametrization can be found 
in the literature.~\cite{SI_slater1929,SI_slater1960,SI_karolakPhD,SI_valliPRR2} 

Under the above conditions, the Coulomb energy difference between the atomic LS and HS states 
in terms of the exchange parameters is given by 
\begin{equation}
 \Delta_{\textrm{Coulomb}} = E^{\textrm{LS}}_{\textrm{Coulomb}}-E^{\textrm{HS}}_{\textrm{Coulomb}} = 2J_1 + 3J_3. 
\end{equation}
The physical interpretation of this result is shown in Fig.~\ref{fig:dCoulomb} and is summarized as follows. 
The LS and HS states differ in the occupation of the \dxy\ and \dxq\ orbitals. 
Since the direct Coulomb interaction is the same for each pair of orbitals, its contribution is the same in both states. 
However, with respect to the LS state, in the HS state there is an additional interaction term $-3J_1$ 
between the parallel \dxy\ and \dxq\ spins 
and two additional terms $-J_1$ since the spin in \dxq\ is now parallel to those in the \dxz\ and \dyz\ orbitals. 
We find that $\Delta_{\textrm{Coulomb}} \sim 2.83$~eV, 
which is of the same order of magnitude as the ligand field splitting 
between the \dxy\ and \dxq\ orbitals. 
Hence, even this approximate picture confirms the competition between the ligand field 
and the Coulomb interaction to be at the root of the SCO phenomenology. 
Moreover, this procedure intuitively shows that the many-body renormalization of the ligand field 
plays a crucial role in the determining which spin state becomes energetically favorable, and at which
critical strain value.

\subsection*{Effective Fe model for transport calculations} 
In the standard quantum transport formalism, 
the transmission function is evaluated through the Landauer formula~\cite{SI_meirPRL68,SI_droghettiPRB95}  
\begin{equation} \label{eq:landauer}
 T(\omega) = Tr \big[ \Gamma^L G^{\dagger} \Gamma^R G \big], 
\end{equation} 
where $G^{(\dagger)}$ denotes the retarded (advanced) real-space Green's function 
of the whole scattering region, 
and $\Gamma^{\alpha} = \imath(\Sigma^{\alpha}-\Sigma^{\dagger\alpha})$ represents the coupling to the leads, 
given in terms of the embedding self-energy $\Sigma^{\alpha}$. 
In general, this approach requires a projection of the whole scattering region 
from plane-waves onto Wannier orbitals, which is far from trivial.~\cite{SI_thygesenCP319,SI_strangeJCP128} 
In particular, for complex devices such as the one considered here, it becomes a prohibitive task. 

Hence, we follow an alternative route, taking advantage of the local projection onto the Fe impurity, 
which was used to evaluate the many-body self-energy. 
We evaluate the Landauer formula of Eq.~(\ref{eq:landauer}) through the Fe-$3d$ orbitals, 
for which the Green's function includes the many-body effects 
stemming from the Coulomb interaction~\cite{SI_meirPRL68,SI_jacob15,SI_droghettiPRB95,SI_valliNL18,SI_valliPRB100} 
and treat the rest of the device as the leads. 
Physically, this corresponds to considering the transmission channels in which the electrons 
tunnel coherently through the correlated Fe atom, 
and the transmission function will qualitatively mirror the spectral properties of the impurity.~\cite{SI_meirPRL68,SI_droghettiPRB95} 
In practice, we define the embedding self-energies  as 
$\Sigma^L(\omega) + \Sigma^R(\omega) = \Delta(\omega)$ 
and, since the device is symmetrical along the transport direction, 
the most natural choice is to assume $\Sigma^L = \Sigma^R$. 
In this framework, the transmission function (per spin) becomes 
\begin{equation}\label{eq:landauer_channels}
 T(\omega) = \sum_{\ell\ell'rr'} \Gamma^L_{\ell\ell'}(\omega) G^{a}_{\ell'r}(\omega) 
                                 \Gamma^R_{rr'}(\omega) G^{r}_{r'\ell}(\omega)
\end{equation}
where indices runs over all orbitals of the Fe-$3d$ multiplet. 

Upon  application of a bias voltage $V_b$, the electric current (per spin) 
driven through the Fe atom is given by
\begin{equation}\label{eq:IV}
 I = \frac{e}{h} \int_{-\infty}^{\infty} d\omega \ T(\omega) \big[ f_L(\omega) - f_R(\omega) \big],
\end{equation}  
where $e$ is the electron charge, $h$ is Plank's constant. 
The Fermi-Dirac distribution function for the $L$ and $R$ electrodes is given by 
\begin{equation}
f_{L/R}(\omega)=\frac{1}{1+\exp[(\omega - \mu_{L/R})/(k_BT)]}, 
\end{equation}
where $\mu_L-\mu_R=V_b$ is the symmetric bias drop, 
$k_B$ is the Boltzmann constant and $T$ the temperature. 
We assumed a temperature broadening $k_BT=25$~meV, 
i.e., close to room temperature, but the results are qualitatively identical  to those obtained for smaller broadenings.
The current in Eq.~(\ref{eq:IV}) is evaluated neglecting the dependence 
of the transmission function on the bias voltage, i.e., $T(E,V_b) \approx T(E)$, 
which is a reasonable assumption in the low-bias regime.~\cite{SI_meirPRL68,SI_nessPRB82,SI_droghettiPRB95}

\bibliographystyle{apsrev}
\bibliography{supplementary}